\documentclass[pra,superscriptaddress,eqsecnum,showpacs]{revtex4}

\begin{document}

\title{Effective Hamiltonians in quantum
optics: a systematic approach}

\author{A. B. Klimov}
\affiliation{Departamento de F\'{\i}sica,
Universidad de Guadalajara,
Revoluci\'on 1500, 44420 Guadalajara,
Jalisco, M\'exico}

\author{L. L. S\'anchez-Soto}
\address{Departamento de \'{O}ptica,
Facultad de Ciencias F\'{\i}sicas,
Universidad Complutense, 28040 Madrid, Spain}

\author{A.Navarro}
\affiliation{Departamento de F\'{\i}sica,
Universidad de Guadalajara,
Revoluci\'on 1500, 44420 Guadalajara,
Jalisco, M\'exico}

\author{E. C. Yustas}
\address{Departamento de \'{O}ptica,
Facultad de Ciencias F\'{\i}sicas,
Universidad Complutense, 28040 Madrid, Spain}

\begin{abstract}
We discuss a general and systematic method for
obtaining effective Hamiltonians that describe different
nonlinear optical processes. The method exploits the
existence of a nonlinear deformation of the usual su(2)
algebra that arises as the dynamical symmetry of the
original model. When some physical parameter, dictated
by the  process under consideration, becomes small, we
immediately get a diagonal effective Hamiltonian
that correctly represents the dynamics for arbitrary
states and long times. We extend the technique to
su(3) and su(N), finding the corresponding effective
Hamiltonians when some resonance conditions are
fulfilled.
\end{abstract}


\maketitle

\section{Introduction}

Quantum optics provides the ideal arena to deal
with the interaction of radiation and matter. Indeed,
by using standard techniques one easily gets a
system of nonlinear coupled equations that govern,
under the approximation of negligible damping, the
interaction of a discrete set of field modes with an
ensemble of atoms~\cite{Cohen98,Walls95,Scully99,Perina91}.
This is a very general description, but too involved
to draw any immediate physical conclusion. Nevertheless,
as soon as one realizes the existence of fast and slow
variables with a large difference in time scales, it is rather
obvious that a simplified formulation can be derived from
first principles, at least for some limiting cases.

In most nonlinear optical phenomena the fields
and the atomic transitions are usually far from
resonance~\cite{Schenzle82}. In fact, no
appreciable population redistribution is brought
about the irradiation even of intense fields, and
only minute fractions of light intensities are
absorbed. In the reference frame of the external
field oscillations, the atomic polarization is then a
fast variable controlled by the slow motion
of the electromagnetic field amplitudes and follows
its evolution adiabatically. Under this assumption we
can eliminate the atomic degrees of freedom and are
left with a small number of equations for the fields
alone. Due to the elimination procedure, these
equations appear as suffering a nonlinear field-field interaction
and can consequently be reinterpreted as the Heisenberg
equations of motion for the field operators under
the dynamics of an effective nonlinear Hamiltonian.
In other words, such an effective Hamiltonian acts only
on the slow degrees of freedom and, because it correctly
describes the slow motion, it incorporates the effect of the
coupling of these slow degrees of freedom with the fast ones.

However, this adiabatic elimination (that was first used
to study the parametric oscillator~\cite{Graham68a,Graham68b})
presents some drawbacks. First, it does not provide a general
prescription for finding effective Hamiltonians, since
the particular details strongly depend on the model
considered. Second, it could become very cumbersome,
and explicit but enormously complicated expressions for
the different orders of approximation can be found in
many original publications~\cite{Bloembergen92,Shen85}.
Third, and even worse, the procedure is not uniquely
defined: depending on the term eliminated, the outcome of the
final Hamiltonian could be different~\cite{Puri88,Gigi90,KC}.

For these reasons, other methods of deriving effective
Hamiltonians exist. For the field of quantum optics, the
ones devised in Refs. \cite{Sczaniecki83} and \cite{Hillery85}
are especially germane. A thorough review of the different ways
of formally constructing effective Hamiltonians may be found in
Ref.~\cite{Klein74}.  Roughly speaking, all the methods
have in common that at some point in their implementation
one applies a unitary (or canonical) transformation to the
total Hamiltonian and keeps only terms up to some fixed
order. It is worth emphasizing that this technique is
standard in condensed matter physics~\cite{Kittel87}
and important examples of its application may be
given~\cite{Shavitt80}.

In consequence, it seems pertinent to find a setting to
support this usual approach of effective Hamiltonians: it is
the main goal of this paper to provide such a setting by
resorting to some elementary notions of group theory.
To this end, we first note that most of the effective
Hamiltonians in quantum optics contain cubic or higher
terms in creation and annihilation operators. Among others,
typical  examples are $k$th harmonic generation, $k$-wave
mixing, and generalized Dicke models. The key point
for our purposes is the recent observation that the common
mathematical structure underlying all these cases is
a nonlinear or polynomial deformation of su(2),
which arises as the dynamical symmetry algebra
of the corresponding Hamiltonian. This nonlinear
algebra has recently found an important place in quantum
optics~\cite{Karasiov92,Karasiov94,Debergh97,Delgado00,Sunil00}
because it allows us to handle the problems in very close
analogy with the usual treatment of an angular momentum.
In particular, we get a decomposition of the Hilbert space
into direct sums of invariant subspaces and the dynamical
problem generated by the corresponding Hamiltonian can be
reduced to the diagonalization of a finite-dimensional matrix.

This is a considerable achievement by itself, but unfortunately
it is impossible to obtain analytic expressions for the
eigenvalues and eigenstates of those matrices. For this
reason,  in Ref.~\cite{Klimov00} we have devised a Lie-like
method~\cite{Steinberg87} that allows one to get approximately
effective Hamiltonians that can be diagonalized in an exact
form.  The idea is to apply a small nonlinear ``rotation'' to the
original Hamiltonian: although, in general, its action is rather
involved, when some physical parameter (dictated by the
particular model under consideration) becomes small, it
generates an effective Hamiltonian that is diagonal and
represents correctly the dynamics for arbitrary states
and even for long times. The method appears then as
the natural and systematic tool for finding effective
Hamiltonians, provided the existence of this nonlinear
su(2) dynamical algebra.

Furthermore, it is worth noting that the su(3) algebra
is the natural extension of su(2) to study the dynamical
evolution of three-level systems. Far from being an exotic
curiosity, this evolution is central to the discussion of
many fascinating problems, such as two-photon
coherence~\cite{Taka75,Loy75}, resonant Raman
scattering~\cite{Stroud76,Cohen77}, superradiance~\cite{BO78}
and three-level echoes~\cite{Mossberg77,Mossberg81}.
As one would expect, when these three-level systems
interact with quantum fields, a nonlinear deformation of
su(3) naturally emerges. Much in the same way, a
su(N) deformed structure arises when considering
the interaction of $N$-level systems with quantized fields.

In this paper we apply our method to some nontrivial
examples: a collection of two-level atoms dispersively interacting
with a quantum field in a cavity; and $\Xi$ and $\Lambda$
configurations of three-level systems interacting with fields
under different resonance conditions. We also obtain effective
Hamiltonians for multilevel systems under $N$-photon resonance
conditions. Finally, we present how our method could work
for more general situations, which would greatly facilitate the
physical applications of this technique to many quantum optical
problems.

\section{Effective Hamiltonians for nonlinear su(2) dynamics}

\subsection{Motivation of the method}

To keep the discussion as self-contained as possible and
to introduce the physical ideas underlying the method, let us
start with the very simple example of a particle of spin $j$
in a magnetic field. The Hamiltonian for this system has the
following form (in units $\hbar = 1$, which will be used
throughout all this paper)
\begin{equation}
\label{Hsu2}
H= \omega S_3 + g ( S_+ + S_- ) ,
\end{equation}
where $g$ is the coupling constant and the operators
$S_3, S_+,$ and $S_-$ constitute a $(2j+1)$-dimensional
representation of the su(2) algebra, obeying the usual
commutation relations
\begin{equation}
\label{ccr}
 [ S_3, S_\pm ] =  \pm S_\pm ,
\qquad
 [ S_+, S_-  ]  =   2S_3 .
\end{equation}
In the traditional angular momentum basis $|j, m \rangle$
($m= -j, -j+1, \ldots, j-1, j$) the operator $S_3$ is diagonal
\begin{equation}
S_3 |j, m \rangle = m |j, m \rangle ,
\end{equation}
while the action of the ladder operators $S_\pm$ is given by
\begin{equation}
S_\pm |j, m \rangle =
\sqrt{(j \mp m) (j \pm m +1 )}
|j, m \pm 1 \rangle .
\end{equation}
The Hamiltonian (\ref{Hsu2}) belongs to the class of the
so-called linear Hamiltonians~\cite{Perinova94} and admits
an exact solution. A convenient way of working out
the solution is to apply the rotation
\begin{equation}
U= \exp \left[ \alpha ( S_+ - S_- ) \right] ,
\label{Usu2}
\end{equation}
and recalling that $e^A B e^{-A} = B + [A, B] +
\frac{1}{2!} [A, [A, B]] + \ldots$, the rotated Hamiltonian,
which is unitarily equivalent to the original one, becomes
\begin{equation}
\label{Halpha}
\tilde{H}  =  UHU^\dagger =
 [ \omega \cos (2\alpha) +2 g \sin (2\alpha) ] S_3 +
\frac{1}{2} [ 2 g \cos(2 \alpha) -
\omega \sin (2\alpha) ] ( S_+ + S_- ) .
\end{equation}

Now, the central idea is to choose the parameter $\alpha$
so as to cancel the nondiagonal terms appearing in
Eq.~(\ref{Halpha}). This can be accomplished by taking
\begin{equation}
\label{aome}
\tan (2\alpha) = \frac{2g}{\omega} ,
\end{equation}
and the transformed Hamiltonian reduces then to
\begin{equation}
\label{Hex}
H_{\mathrm{eff}}= \omega
\sqrt{1+ \frac{4g^2}{\omega ^2}} \
S_3 .
\end{equation}
Since this effective Hamiltonian is diagonal in the angular
momentum basis, the dynamical problem is completely solved.
The crucial observation  is that when $ g \ll \omega$ we
can approximate Eq.~(\ref{aome}) by $\alpha \simeq
g/\omega$, and then Eq.~(\ref{Usu2}) can be substituted by
\begin{equation}
U \simeq \exp \left[ \frac{g}{\omega }
( S_+ - S_- ) \right]  .
\end{equation}
This small rotation approximately (i.e., up to second-order terms
in $g/\omega $) diagonalizes the Hamiltonian (\ref{Hsu2}),
giving rise to the effective Hamiltonian
\begin{equation}
\label{Hefect}
H_{\mathrm{eff}}= U H U^\dagger
\simeq
\left ( \omega +2 \frac{g^2}{\omega} \right)  S_3
\end{equation}
which obviously coincides with the exact solution~(\ref{Hex})
after expanding  in a series of $g^2/\omega^2$.
A direct application of the standard time-independent
perturbation theory~\cite{Cohen90} to Eq.~(\ref{Hsu2})
leads immediately to the same eigenvalues and eigenstates
that  the Hamiltonian~(\ref{Hefect}) gives in the same
order of approximation. However, we stress that our method
is fully operatorial and avoids the tedious work of computing
the successive corrections as sums over all the accessible states.

\subsection{Nonlinear small rotations}

Having in mind the previous simple example, let us
go one step further by treating the more general case
of a system that admits some integrals of motion $N_j$
and whose interaction Hamiltonian can be written as
\begin{equation}
 \label{Hint}
H_{\mathrm{int}} = \Delta \ X_3+g (X_+ + X_-) ,
\end{equation}
where $g$ is a coupling constant and $\Delta $ is a parameter
usually representing the detuning between frequencies of
different  subsystems (although it is not necessary). The
operators $X_\pm $ and $X_3$ maintain the first commutation
relation of su(2), $[X_3, X_\pm ] = \pm X_\pm,$ but the
second one is modified in the following way
\begin{equation}
\label{P}
[X_+,  X_-] = P(X_3) ,
\end{equation}
where $P(X_3)$ is an arbitrary polynomial function
of the  diagonal operator $X_3$ with coefficients
perhaps depending  on the integrals of motion $N_j$.
These commutation relations correspond to the
so-called polynomial deformation of su(2). Such
nonlinear algebras were discovered by
Sklyanin~\cite{Skl82} and Higgs~\cite{Higgs79}
and have already played an  important  role in quantum
mechanics~\cite{Rocek91,Bonatsos93,Quesne95,Beckers96}.

Now suppose that for some physical reasons (depending
on the particular model under  consideration) the condition
\begin{equation}
\label{epsilon}
\varepsilon =\frac{g}{\Delta }\ll 1
\end{equation}
is fulfilled. Then, it is clear that (\ref{Hint}) is {\it almost}
diagonal in the basis that diagonalizes $X_3$. In fact, a
standard perturbation analysis immediately shows that the
first-order corrections introduced  by the nondiagonal part
$g(X_+ + X_-)$ to the eigenvalues of $X_3$ vanish and
those of second order are proportional to $\varepsilon$.
Thus, we apply the following unitary transformation to
(\ref{Hint})  (which, in fact, is a \textit{small} nonlinear
rotation)
\begin{equation}
\label{small}
U= \exp [ \varepsilon  (X_+-X_-) ] .
\end{equation}
After some calculations we get, up to order $\varepsilon^2$,
the effective Hamiltonian we are looking for
\begin{equation}
H_{\mathrm{eff}} = \Delta \ X_3 +
\frac{g^2}{\Delta} P(X_3),
\label{H1eff}
\end{equation}
which is diagonal in the basis of eigenstates of  $X_3$.
Then,  the evolution (as well as the spectral) problem is
completely  solved in this approximation. Besides the
advantage of having  the effective Hamiltonian expressed
in an operatorial form,  the method has the virtue of
generality, since it is valid for any model whose
Hamiltonian could be written down in terms of the
generators of a polynomial deformation of su(2).

Our technique also provides a valuable tool for obtaining
corrections to the eigenstates of (\ref{Hint}). Indeed, it
is easy to realize that these eigenstates can be
approximated by
\begin{equation}
| \Psi_m \rangle = U^{\dagger }| m \rangle ,
\end{equation}
where $|m\rangle $ denotes an eigenstate of  $X_3$ and
$U$ is the corresponding small rotation. Since $U$ and
$|m\rangle $  do not depend on time, the operator $U$
can be applied to $|m\rangle $ as an expansion in
$\varepsilon$. For example,  the eigenstate
$|\Psi_m \rangle $ up to order $\varepsilon^2$
takes on the form
\begin{equation}
| \Psi_m \rangle = \left [ 1- \varepsilon \ (X_+ - X_-)
- \frac{\varepsilon^2}{2} \
(1 + 2 X_+ X_- - X_+^2 - X_-^2 )
 \right  ] | m \rangle.
\end{equation}
This representation is especially advantageous when the state space
of the model is a representation space of the deformed su(2)
algebra that is constructed in the usual way by the action of  the
raising operator $X_+$; i.e., $ | m \rangle \sim X_+^{m} | 0 \rangle,$
where $|0 \rangle $ is a lowest weight vector fulfilling the standard
condition $X_- | 0 \rangle =0$.

This general procedure shows that can always adiabatically eliminate
all nonresonant transitions and work with an effective Hamiltonian
containing only (quasi) resonant  transitions. The effect of nonresonant
terms reduces to a dynamical Stark shift (which can have a quite
complicated form). Obviously, transformations generating effective
Hamiltonians also change eigenfunctions, but the corrections are of
order $\varepsilon$ and do not depend on time. Such corrections
correspond to low-amplitude transitions that take place in the case
of nonresonant interactions. We shall elaborate on these topics
in the next Sections.

\subsection{An example: dispersive limit of the Dicke model}

As a relevant example, let us apply our method to the well-known
Dicke model that describes the interaction of a single-mode
field  of frequency $\omega_{\mathrm{f}}$
with a collection  of $A$ identical  two-level atoms with
transition frequency $\omega_0$. Making the standard
dipole and rotating-wave approximations, the model
Hamiltonian reads as~\cite{Dicke54}
\begin{equation}
\label{Dicke1}
H =  \omega_{\mathrm{f}}  a^\dagger a +
\omega_0 S_3 + g ( aS_+ + a^\dagger S_- ) ,
\end{equation}
where $a (a^\dagger)$ are the annihilation (creation)
operators for the field mode, and $(S_\pm , S_3)$ are
collective atomic operators forming an
$(A+1)$-dimensional representation of su(2).

By introducing the excitation number $N = a^\dagger a + S_3$,
which is an integral of motion, we can recast (\ref{Dicke1})
as $H= H_0 + H_{\mathrm{int}}$  with
\begin{eqnarray}
\label{HD}
H_0 &=& \omega_{\mathrm{f}} N ,
\nonumber \\
& & \\
H_{\mathrm{int}} & = &
\Delta \ S_3 + g ( aS_+ + a^\dagger  S_- )  ,
\nonumber
\end{eqnarray}
the detuning being $ \Delta = \omega_0 -
\omega_{\mathrm{f}}.$

We assume now that the dispersive limit
holds~\cite{Brune96}; i.e.,
\begin{equation}
| \Delta | \gg A g \sqrt{\bar{n}+1} ,
\end{equation}
where $\bar{n}$ is the average number of photons in
the field.  If, after our previous discussion, we
introduce the deformed su(2) operators as
\begin{equation}
X_+ = a S_+, \qquad
X_- = a^\dagger S_-,
\qquad
X_3 = S_3,
\end{equation}
then we immediately get that the small nonlinear
rotation (\ref{small}) transforms the interaction
Hamiltonian~(\ref{HD}) into
\begin{equation}
\label{H2}
H_{\mathrm{eff}} =   \Delta \ S_3 +
\frac{g^2}{\Delta} [ S_3^2 -
2 (a^\dagger a +1 ) S_3 - C_2 ] ,
\end{equation}
where $C_2 = A/2 ( A/2+1) $ is the value of the Casimir
operator for su(2). The effective Hamiltonian (\ref{H2})
was previously obtained in Ref.~\cite{Agarwal97} by quite a
different method (see also Ref.~\cite{Klimov98}) and,
due to the presence of the nonlinear term $S_3^2$, has been
considered as a candidate for the generation of squeezed
atomic states~\cite{Ueda93}.

\section{Effective Hamiltonians for nonlinear su(3) dynamics}

\subsection{Three-level systems interacting with quantum fields}

The method of small rotations discussed in the previous
Section can be applied not only to Hamiltonians having a
nonlinear su(2) structure, but also to more complicated
systems.  In this Section we shall focus on Hamiltonians
that can be represented in terms of  the su(3) algebra.
This structure naturally arises when dealing with systems
with three relevant levels. It is well known that in this
case three possible configurations (commonly  called
$\Xi ,V,$ and $\Lambda $) are admissible
(see  Ref.~\cite{YE85} for details).

For definiteness, we shall consider the interaction of $A$
identical three-level systems in a cascade or $\Xi $ configuration
(i.e., with associated energies $E_1 < E_2 < E_3$ and
allowed dipole transitions $1 \leftrightarrow 2$ and
$2 \leftrightarrow 3$, but not the $1 \leftrightarrow 3$)
interacting with a single-mode quantum field of frequency
$\omega_{\mathrm{f}} $. The Hamiltonian of this model is
\begin{equation}
\label{H3xi}
H_\Xi = H_{\mathrm{field}} + H_{\mathrm{atom}}
+ H_{\mathrm{int}} ,
\end{equation}
with
\begin{eqnarray}
\label{H3s}
H_{\mathrm{field}} & = &  \omega_{\mathrm{f}} a^\dagger a ,
\nonumber \\
H_{\mathrm{atom}} & = & E_1 S^{11}+
E_2S^{22}+E_3S^{33},  \\
H_{\mathrm{int}}&  =  &
g_{12} (a S_+^{12} + a^\dagger S_-^{12}) +
g_{23} (a S_+^{23} + a^\dagger S_-^{23}) .
\nonumber
\end{eqnarray}
As usual, the three diagonal observables $S^{ii}= |i \rangle \langle i |$
($i=1,2,3$) measure the population of the $i$th energy level, while
the off-diagonal polarizations $S^{ij}= | j \rangle \langle i | $ generate
transitions from level $i$ to $j$. They satisfy the commutation relations
\begin{equation}
\label{ccru3}
[S^{ij}, S^{kl}] = \delta_{jk} S^{il} - \delta_{il} S^{kj} ,
\end{equation}
which turn out to be those of the u(3) algebra. In fact, they
form a $(A+1) (A+2)/2$-dimensional representation of the
u(3) algebra.

Because the sum $S^{11}+S^{22}+S^{33}=A $ is an
obvious integral of motion that determines the total number of atoms,
only two of the populations can vary independently. For this
reason it is customary to introduce the two traceless operators
\begin{equation}
S_3^{12} = \frac{1}{2} ( S^{22} - S^{11}) ,
\qquad
S_3^{23} = \frac{1}{2} (S^{33}-S^{22}) ,
\end{equation}
that represent population inversion between the corresponding
levels. Furthermore, to emphasize the idea of transition between
levels, it is usual to define the operators $S_+^{ij} = S^{ij}$
and $S_-^{ij} = S^{ji}$ (for $i<j$). Then the eight independent
operators $(S_\pm^{ij}, S_{3}^{ij})$ satisfy the commutation
relations of su(3). Moreover,  $(S_\pm^{12}, S_3^{12})$ and
$(S_\pm^{23}, S_3^{23})$ form two su(2) subalgebras (each
one  describing the transitions $1\leftrightarrow 2$ and
$2\leftrightarrow 3$). Nevertheless, transitions $1\leftrightarrow 2$
and  $2\leftrightarrow 3$ are not physically independent, since
we have
\begin{equation}
\label{Scr1}
[S_+^{12}, S_+^{23}] = -S_+^{13} ,
\qquad
[S_-^{12}, S_-^{23}] = S_-^{13},
\qquad
[S_+^{12}, S_-^{23}] = 0 .
\end{equation}

It seems natural also to introduce the deformed su(3) algebra as
\begin{eqnarray}
\label{DG}
& X^{11}= S^{11}, \qquad
X^{22}= S^{22}, \qquad
X^{33}=S^{33}, & \nonumber \\
& & \\
& X_+^{12} = a S_+^{12},
\qquad
X_+^{23}=aS_+^{23} .  &
\nonumber
\end{eqnarray}
Then the Hamiltonian (\ref{H3xi}) can be recast as
$H_\Xi = H_0 + H_{\mathrm{int}}$, with
\begin{eqnarray}
H_0 & = & \omega_{\mathrm{f}} N_\Xi , \nonumber \\
& & \\
H_{\mathrm{int}} & = & - \Delta _{12} \ X^{11}+
\Delta_{23} \ X^{33}+
g_{12} ( X_+^{12} + X_-^{12}) +
g_{23} ( X_+^{23} + X_-^{23}) , \nonumber
\end{eqnarray}
where we have used the conserved excitation number
$N_\Xi = a^\dagger a+ S^{33} - S^{11}$ and we have
introduced the detunings as
\begin{equation}
\Delta_{12} = E_2 - E_1 - \omega_{\mathrm{f}} ,
\qquad
\Delta_{23} = E_3 - E_2 - \omega_{\mathrm{f}} .
\end{equation}
The operators $X^{ij}$ satisfy the usual su(3)
commutation relations, with (\ref{Scr1}) recast as
\begin{equation}
\label{su3}
[X_+^{12}, X_+^{23}] = -Y_+^{13} ,
\qquad
[X_-^{12}, X_-^{23}] = Y_-^{13},
\qquad
[X_+^{12}, X_-^{23}] = 0 .
\end{equation}
where
\begin{equation}
\label{Y}
Y_+^{13} = a^2 S_+^{13} .
\end{equation}
However, we have to modify some of them in the following way:
\begin{equation}
 \label{Sdcr3}
[X_+^{ij}, X_-^{ij}] = P(X^{ii}, X^{jj}) ,
\qquad
[Y_+^{ij},Y_-^{ij}] = Q(X^{ii}, X^{jj}) ,
\end{equation}
where $P(X^{ii},X^{kk})$ and $Q(X^{ii},X^{kk})$ are
polynomials of the diagonal operators $X^{ii}$ ($i=1,2,3$)
and  define, then, a polynomial deformation of su(3).

\subsubsection{Effect of a far-off resonant level}

The dynamics generated by su(3) is obviously richer
than that of su(2), since a greater number of physical
degrees of freedom are now available. To see how our
method works in this case, let us assume that one of
the transitions, say the $2\leftrightarrow 3$, is (quasi)
resonant with the field; i.e.,
\begin{equation}
| \Delta _{23} | \ll  A g_{23}\sqrt{\bar{n}+1},  \label{rcL1}
\end{equation}
while the transition $1\leftrightarrow 2$ is far-off resonant
\begin{equation}
|\Delta _{12}| \gg A g_{12} \sqrt{\bar{n}+1} .  \label{dcL1}
\end{equation}
It is clear from our previous analysis that  the
small nonlinear rotation
\begin{equation}
\label{U12}
U_{12} = \exp [ \varepsilon_{12} (X_+^{12} - X_-^{12}) ] ,
\end{equation}
with $\varepsilon_{12} = g_{12}/\Delta_{12} \ll 1$, eliminates
the interaction term $g_{12}(X_+^{12}+X_-^{12})$, representing
the nonresonant transition $1\leftrightarrow 2$. In such a way, we
obtain  the effective Hamiltonian
\begin{eqnarray}
\label{HL1}
H_{\mathrm{eff}} & = & \Delta _{12}\ X^{11}+
\Delta _{23}\ X^{33}+g_{23} ( X_+^{23} + X_-^{23}) -
\frac{g_{12}g_{23}}{\Delta _{12}} (Y_+^{13}+Y_-^{13})
\nonumber \\
& + &  \frac{g_{12}^2}{\Delta_{12}} P(X^{11}, X^{22})
+\frac{g_{12}g_{23}}{\Delta _{12}}
( [X_+^{12}, X_-^{23}] + [X_+^{23}, X_-^{12}] ) .
\end{eqnarray}
The remarkable point is that by eliminating the transition
$1\leftrightarrow 2$, we have generated an effective transition
$1\leftrightarrow 3$ (represented by the operators
$Y_{\pm }^{13}$),  which was absent in the initial Hamiltonian.
Nevertheless, this transition is also nonresonant due to
conditions (\ref {rcL1}) and (\ref{dcL1}) and, accordingly,
can be eliminated by the following transformation
\begin{equation}
U_{13}=\exp [ \varepsilon_{13} (Y_+^{13}-Y_-^{13}) ] ,
\label{U3}
\end{equation}
where the parameter $\varepsilon_{13} $ in the above equation
must fulfill
\begin{equation}
\varepsilon_{13} = \frac{g_{12} g_{23}}
{\Delta _{12}(\Delta _{12}+\Delta _{23})}\ll 1.
\end{equation}

In this particular case the polynomial function $P(X^{ii},X^{jj})$ is
\begin{equation}
P(X^{11}, X^{22})=  S_+^{12} S_-^{12} +
a^\dagger a (S^{22}-S^{11}).
\label{Pg}
\end{equation}
If initially level 1 is unpopulated we have $S^{11}=0$
(which will be conserved, since there are no transitions to
the level 1), and
\begin{equation}
P(X^{11}, X^{22}) = S^{22}(a^\dagger a +1).
\end{equation}
In consequence,  the effective Hamiltonian, taking into
account the existence of a far-off resonant level, has
the form
\begin{equation}
H_{\mathrm{eff}} = \Delta _{23} \ S^{33}+
g_{23} (a S_+^{23}+ a^\dagger S_-^{23}) +
\frac{g_{12}^2}{\Delta _{12}}S^{22}(a^\dagger a + 1).
\end{equation}
This means that the far-lying level produces a mark in the system
in  the form of a dynamical Stark-shift term. It is interesting to
observe that due to this Stark shift, the initially nonresonant
transition  $2\leftrightarrow 3$ becomes resonant in a
subspace with some fixed photon number. This opens the
possibility of separating $n$-photon field states from an  initial
coherent state interacting with a collection of three-level atoms
just by choosing a suitable relation between the detunings and
the interaction constants, and projecting to the second level
at appropriate moments.

\subsubsection{Two-photon resonance}

Let us now envisage the different situation in which the
two-photon resonance condition between levels 1 and 3 is fulfilled;
i.e., $E_3-E_1 = 2 \omega_{\mathrm{f}}$. This means that $\Delta _{12}
= - \Delta _{23}$ and the transition generated by the operators
$Y_{\pm }^{13}$ cannot be removed. However, the term
$g_{23}(X_+^{23}+X_-^{23})$, which generates (nonresonant)
transitions between levels 2 and 3, can be eliminated by a
transformation $U_{23}$ analogous to (\ref{U12}) with
$\varepsilon_{23}\ll 1.$ The transformed Hamiltonian
becomes then
\begin{eqnarray}
H_{\mathrm{eff}} & = & - \Delta _{12}\ ( X^{11}+X^{33}) -
\frac{g_{12}g_{23}}{\Delta _{12}} (Y_+^{13}+Y_-^{13})
\nonumber \\
& + & \frac{g_{12}^2}{\Delta _{12}}P(X^{11}, X^{22})
- \frac{g_{23}^2}{\Delta_{12}} P(X^{22}, X^{33}),
\end{eqnarray}
where $P(X^{11},X^{22})$ and $P(X^{22},X^{33})$ are
defined according to  Eq.~(\ref{Pg}). Finally, if we further
impose the  absence of initial population in level 2, we obtain
\begin{eqnarray}
\label{H2ph}
H_{\mathrm{eff}} & = & \frac{g_{12}g_{23}}
{\Delta _{12}}(a^2S_+^{13}+ {a^\dagger}^2 S_-^{13})
\nonumber \\
& +& ( S_3^{13} + A/2 ) \left [  ( g_{23}^2/\Delta _{12} -
g_{12}^2/\Delta_{12} ) a^\dagger a + g_{23}^2/\Delta _{12}
\right ] + A \frac{g_{12}^2}{\Delta _{12}} a^\dagger a,
\end{eqnarray}
which is the effective two-photon Dicke Hamiltonian including the
dynamical Stark shift obtained by Puri and Bullough~\cite{Puri88}
(see also Ref.{\cite{KC}).

\subsection{The dispersive limit of the $\Lambda$ configuration}

Let us consider for a moment the case of the  $\Lambda $
configuration,  in which the allowed dipole transitions are
now $1\leftrightarrow 3$ and $2\leftrightarrow 3$, but not
the $1\leftrightarrow 2$. The Hamiltonian governing the
evolution  is still of the form (\ref{H3xi}), but now with
\begin{equation}
H_{\mathrm{int}} =
g_{13} ( a S_+^{13}+ a^\dagger S_-^{13}) +
g_{23} (a S_+^{23}+a^\dagger S_-^{23}) .
\end{equation}
By using the integral of motion $N_\Lambda =
a^\dagger a + S^{33}$ we can rewrite
\begin{equation}
\label{HLi}
H_{\mathrm{int}} = - \Delta _{31}X^{11}-
\Delta _{32}X^{22} + g_{13} (X_+^{13} + X_-^{13} )
+  g_{23} ( X_+^{23} + X_-^{23} ) ,
\end{equation}
where, as in Eq.~(\ref{DG}),  we have introduced the
deformed  su(3) operators as
\begin{equation}
\label{DG1}
X_+^{13} = a S_+^{13},
\qquad
X_+^{23}=aS_+^{23} .
\end{equation}
These deformed generators satisfy a set of commutation
relations similar to (\ref{su3}), but instead of (\ref{Y})
we must use
\begin{equation}
Y_+^{12} = ( S^{33} - a^\dagger a ) ,
\end{equation}
which is the mathematical reason for the well-known
different behaviours exhibited by $\Xi$ and $\Lambda$
configurations.

Let us focus on the dispersive regime, when
\begin{equation}
|\Delta _{13}| \gg A g_{13} \sqrt{\bar{n}+1},
\qquad
 |\Delta _{23}| \gg Ag_{23} \sqrt{\bar{n}+1} .
\end{equation}
Then a couple of small rotations eliminate the
far-off resonant transitions $1\leftrightarrow 3$ and
$2\leftrightarrow 3$,  obtaining
\begin{eqnarray}
H_{\mathrm{eff}} & = &   -\Delta_{31}S^{11} -
\Delta _{32}S^{22} \nonumber  \\
& + &\frac{g_{13}^2}{\Delta_{13}}
[ ( S^{11} + 1 ) S^{33} + a^\dagger a ( S^{33} - S^{11}) ]
+ \frac{g_{23}^2}{\Delta_{23}}
[ ( S^{22} + 1 ) S^{33} + a^\dagger a (S^{33} - S^{22} ) ]
\nonumber \\
&+ & \frac{g_{13} g_{23}}{\Delta_{31}}
( S_+^{12} + S_-^{12}) ( S^{33}-a^\dagger a ) .
\end{eqnarray}
Note that both $a^\dagger a$ and $S^{33}$ are now integrals
of  motion. The two first terms correspond to trivial free atomic
dynamics. The next two terms represent the standard
dynamical Stark shift. Finally, the last term describes an
effective interaction between levels 1 and 2. The remarkable
point is that there is a population transfer (and not only phase
transfer, as it could be expected from a dispersive interaction)
between these two levels without exchange of photons. The
intensity of  the transition $1\leftrightarrow 2$ depends on the
difference between the population of level 3 and the photon
number.  Thus, no population transfer between levels 1 and 2
will occur in the sector where the number of photons is
exactly equal to the initial population of the level 3. It is
easy to observe that the (effective) transitions $1\leftrightarrow 2$
are stronger when $\Delta _{31}=\Delta _{32}$; i.e., when
the levels 1 and 2 have the same energy (Zeeman-like systems).

\section{Effective Hamiltonians for nonlinear su(N) dynamics}

\subsection{Multilevel systems interacting with a quantum field}

Having demonstrated the role played by nonlinear algebras
in the systematic construction of effective Hamiltonians,
we would like to pursue here a natural extension to
multilevel systems.

Specifically,  we are interested in considering Hamiltonians that
can be represented in terms of the su(N) algebra, which naturally
arises when describing systems with $N$ relevant levels.
Obviously,  the evolution of a collection of $A$ identical $N$-level
systems (for definiteness, we assume a cascade configuration,
such that $E_i < E_j$ for $i<j$) interacting with a single-mode
quantum field can be modeled by a Hamiltonian as in
Eq.~(\ref{H3xi}) with
\begin{eqnarray}
\label{H}
H_{\mathrm{field}} & = &  \omega_{\mathrm{f}} a^\dagger a ,
\nonumber \\
H_{\mathrm{atom}} & = & \sum_{i=1}^{N}E_i S^{ii}  , \\
H_{\mathrm{int}}&  =  &
\sum_{i=1}^{N-1}g_i (a S_+^{ii+1}+
a^\dagger S_-^{ii+1}) ,  \nonumber
\end{eqnarray}
where $S^{ii}$ ($i=1, \ldots, N$) are population operators of
the $i$th energy level, and $S_\pm^{ij}$ describe transitions
between levels $i$ and $j$. The operators $S^{ij}$ form the
u(N) algebra and satisfy the commutation relations
\begin{equation}
[ S^{ij}, S^{kl} ] = \delta _{jk} S^{il} - \delta _{il} S^{kj}.
\end{equation}
By introducing inversion-like operators
\begin{equation}
S_3^{i i+1} = \frac{1}{2} (S^{i+1 i+1} - S^{ii} ) ,
\end{equation}
then  $(S_{\pm }^{ij}, S_3^{i i+1})$ turn out to be the su(N)
algebra.

The Hamiltonian (\ref{H}) admits the integral of motion
\begin{equation}
\label{IM}
N = a^\dagger a + \sum_{i=1}^{N-1} \mu_i S_3^{i i+1},
\end{equation}
with $\mu_i = i (N-i)$. We also introduce the detunings by
\begin{equation}
\Delta_j = E_j - E_1- ( j-1 ) \omega_{\mathrm{f}} ,
\end{equation}
and we shall assume that $\Delta_j$ satisfy the following resonant
condition
\begin{equation}
\label{rc}
\Delta_N = 0 ,
\end{equation}
which means that the field in a ($N-1$)-photon resonance with
the atomic system; i.e., $E_N - E_1= (N-1)\omega_{\mathrm{f}}$. Thus,
the Hamiltonian (\ref{H}) can be recast as $ H = H_0 +
H_{\mathrm{int}}$, with
\begin{eqnarray}
\label{recast}
H_0 & = & \omega_{\mathrm{f}} N +E A \nonumber \\
& & \\
H_{\mathrm{int}} & = & h_0 + V,  \nonumber
\end{eqnarray}
where $E = (E_N + E_1)/2$ and
\begin{equation}
h_{0}  =  \sum_{i=1}^N
\Delta_i S^{ii},
\qquad
V  =  \sum_{i=1}^{N-1} g_i (a S_+^{ii+1}+a^\dagger S_-^{ii+1}) .
\end{equation}
From our previous experience it seems rather obvious that
the operators
\begin{equation}
X^{ii} = S^{ii},
\qquad
X_+^{ij}= a S_+^{ij},
\qquad
X_-^{ij}= a^\dagger S_-^{ij} ,
\end{equation}
form a polynomial deformation of su(N). In consequence,
and according to our general scheme,  we can introduce the
transformation
\begin{equation}
\label{T1}
U= \exp \left[ \sum_{i=1}^{N-1} \varepsilon_i
(X_+^{i i+1} - X_-^{i i+1})  \right ] ,
\end{equation}
where
\begin{equation}
\varepsilon_i = \frac{g_i}{\Delta_{i+1}- \Delta_i}
\label{alfa}
\end{equation}
will be supposed to be small numbers, $\varepsilon_i \ll 1$,
which means that the transitions are far from the  one-photon
resonance ($\Delta_{i+1}-\Delta_i = E_{i+1}-E_i -
\omega_{\mathrm{f}} \gg g_i$). Thus,  all one-photon
transitions are eliminated by (\ref{T1}) and the
transformed Hamiltonian takes the form
\begin{equation}
H_{\mathrm{eff}}^{(1)} =
h_0 +h_{\mathrm{d}} +
h_{\mathrm{nd}}
+ \sum_{n=1}^{N-2} \frac{n}{(n+1)!}
\sum_{i=1}^{N-n-1} \lambda_i^{(n+1)}
\left ( a^{n+1} S_+^{i  i+n+1} +
{a^\dagger}^{n+1} S_-^{i i+n+1}
\right )  .  \label{He}
\end{equation}
Here, the effective interaction constants $\lambda_i^{n}$
can  be obtained from the recurrence relation
\begin{equation}
\lambda_i^{(n+1)}= \varepsilon_{i+n} \lambda_i^{(n)} -
\varepsilon_i \lambda_{i+1}^{(n)},  \label{psi}
\end{equation}
with the initial term $\lambda_i^{(1)} = g_i$. It is easy to
see that $\lambda_i^{(n+1)}\ll \lambda_i^{(n)}$.

The piece $h_{\mathrm{d}}$ contains only diagonal terms
in the atomic operators and depends on the integral of motion
$N$ (or, equivalently, depends only on the photon-number
operator $a^\dagger a$). This operator $h_{\mathrm{d}}$
appears naturally represented as an expansion in the small
parameter $\varepsilon_i$ whose first term is
\begin{equation}
\label{diag}
h_{\mathrm{d}} = \sum_{i=1}^{N-1} g_i \varepsilon_i
[ a^\dagger a  (S^{i+1 i+1}- S^{ii}) +
(S^{ii} + 1 ) S^{i+1 i+1} ] .
\end{equation}
The essential point for our purposes is that, given its structure,
this diagonal part cannot be removed from the effective
Hamiltonian (\ref{He}). On the contrary, the operator
$h_{\mathrm{nd}}$ contains only nondiagonal terms that
can be eliminated  by rotations of the type (\ref{T1})
unless some specific resonance conditions are fulfilled.
In this respect, let us note that the price we pay for
eliminating one-photon transitions is the generation of all
possible $k$-photon transitions ($k=2,...,N-1$).

\subsection{Three-photon resonance}

Let us consider the particular case of four-level systems
($N=4$) and suppose that there are no transitions in
one- and two-photon resonance with the field. After
eliminating one-photon transitions the transformed
Hamiltonian (\ref{He}) has the form
\begin{equation}
 \label{H3}
H_{\mathrm{eff}}^{(1)}  =  h_0 +
h_{\mathrm{d}} + h_{\mathrm{nd}}
+\frac{1}{3} \lambda_1^{(3)}
\left ( a^3 S_+^{14}+ {a^\dagger}^3 S_-^{14} \right )
+ \frac{1}{2} \sum_{i=1}^{2} \lambda_i^{(2)}
\left (a^2 S_+^{i i+2} + {a^\dagger}^2 S_-^{i i+2} \right ) ,
\end{equation}
where the interaction constants are defined, according to
(\ref{alfa}) and  (\ref{psi}), as
\begin{eqnarray}
\label{lacou}
& \displaystyle \varepsilon_1 = \frac{g_1}{\Delta _2} ,
\qquad
\varepsilon_2 = \frac{g_2}{\Delta _3 - \Delta _2},
\qquad
\varepsilon_3 = - \frac{g_3}{\Delta _3} , & \nonumber \\
& & \\
& \displaystyle
\lambda_1^{(2)} = g_1 g_2 \frac{2\Delta_2-\Delta_3}
{\Delta_2 ( \Delta _3 - \Delta _2)},
\qquad
\lambda_2^{(2)} = g_2 g_3 \frac{2 \Delta_3 - \Delta_2}
{\Delta _3 ( \Delta _2 - \Delta _3) },
\qquad
\lambda_1^{(3)} = \frac{3g_1g_2g_3}{\Delta _3\Delta _2} ,
\nonumber
\end{eqnarray}
and the resonance condition  $\Delta _{4}=0$ (that is,
three-photon resonance $E_{4}-E_1= 3 \omega_{\mathrm{f}}$)
has been imposed.

According to the general scheme, the term representing
two-photon  transitions  in (\ref{H3}) can be removed
using a transformation  analogous to (\ref{T1}):
\begin{equation}
\label{U21}
U_2 = \exp \left [ \frac{1}{2} \sum_{i=1}^2
\alpha_i^{(2)} (a^2 S_+^{i i+2} -
{a^\dagger}^2 S_-^{i i+2}) \right ] ,
\end{equation}
where
\begin{equation}
\alpha_i^{(2)} = \frac{\lambda_i^2}{\Delta_{i+2}-\Delta_i}
\end{equation}
is a small parameter because there are no resonant
two-photon transitions ($\Delta_{i+2} - \Delta_i =
E_{i+2}-E_i - 2 \omega_{\mathrm{f}} \gg \lambda_i^{(2)}$) and thus
$\alpha _{j}^{(2)}\ll \varepsilon_{j}$. It is worth noting
that the transformation (\ref{U21}) does not introduce
new terms of order $\varepsilon^2$ to the effective
Hamiltonian.

The diagonal part in (\ref{H3}) is
\begin{equation}
h_{\mathrm{d}} = \sum_{i=1}^3
g_i \varepsilon_i \left [ a^\dagger a (S^{i+1 i+1} - S^{ii} )
+ (S^{ii}+1) S^{i+1 i+1} \right]  ,
\end{equation}
while the nondiagonal term  deserves a more careful
analysis. Its explicit form is
\begin{equation}
h_{\mathrm{nd}} = \frac{1}{2}
( \varepsilon_1 g_3 + \varepsilon_3 g_1)
(S_+^{12} S_-^{34} + S_+^{34} S_-^{12} ) +
\frac{1}{2}
( \varepsilon_1 g_2 + \varepsilon_2 g_1 )
( S_+^{12} S_-^{23} + S_+^{23} S_-^{12} ) .
\end{equation}
It is clear that the first term in the above expression
describes a resonant dipole-dipole interaction, under
the condition $\Delta _2=-\Delta _3$. On the other
hand, the second term describes a resonant interaction
whenever $ 2 \Delta_2 = \Delta_3$, which is incompatible
with the pervious one and the absence of one- and two-photon
resonances. If  no one of these conditions are fulfilled, the term
$h_{\mathrm{nd}}$ can be eliminated by the transformation
\begin{equation}
U_2^{(2)}= \exp \left [\frac{1}{2}
\sum_{i,j=1}^{3} \beta_{ij}
(S_+^{ii+1}S_-^{jj+1}-S_+^{jj+1}S_-^{ii+1})
\right] ,
\end{equation}
where
\begin{equation}
\beta_{ij} = \frac{\varepsilon_i g_j}{\Delta _{i+1}-\Delta _{i}+
\Delta_j - \Delta _{j+1}}.
\end{equation}
Then, since $S^{11}+S^{22}+S^{33}+S^{44}=A$
and imposing the condition of the absence of initial
population in levels 2 and 3, we obtain the effective
Hamiltonian describing three-photon resonant transitions
\begin{equation}
\label{H3f1}
H^{(2)}_{\mathrm{eff}} =
\frac{g_1g_2g_3}{\Delta _2\Delta _3}
(a^3 S_+^{14} + {a^\dagger}^3 S_-^{14} ) -
( S_3^{14} + A/2 ) [ a^\dagger a
(g_1^2/\Delta _2 - g_3^2 / \Delta _3 ) +
g_3^2/\Delta _3 ] + A \frac{g_1^2}{\Delta _2} a^\dagger a .
\end{equation}

The effective three-photon Hamiltonian (\ref{H3f1}) contains
a dynamical Stark shift similar to that appearing in the two-photon
case (\ref{H2ph}). Nevertheless, the essential difference
consists in that in the two-photon case the interaction term
and the Stark shift are of the same order of magnitude, while
in the three-photon case the interaction term is one order of
magnitude less than the Stark shift. This would lead to essential
differences in the evolution of some observables.

\subsection{Multimode fields and resonance conditions}

The formalism developed can also be used to treat the
interaction of multifrequency fields with atomic systems.
This type of interaction is much richer because some
additional resonance conditions can be satisfied. To this
end, let us consider a four-level system in a $\Xi$ configuration
interacting with two field modes of frequencies $\omega_a$ and
$\omega_b$, and annihilation operators $a$ and $b$, respectively.
The Hamiltonian describing this interaction is still of
the form (\ref{H3xi}) with
\begin{eqnarray}
\label{H4}
H_{\mathrm{field}} & = &  \omega_a a^\dagger a +
\omega_b b^\dagger b , \nonumber \\
H_{\mathrm{atom}} & = & \sum_{i=1}^{4}E_i S^{ii} ,  \\
H_{\mathrm{int}}&  =  &
\sum_{i=1}^{3} \left [ ( g_{ai} a + g_{bi} b) S_+^{i i+1} +
( g_{ai} a^\dagger  + g_{bi} b^\dagger ) S_-^{i i+1} \right ] ,
\nonumber
\end{eqnarray}
where $g_{ai}$ and $g_{bi}$ ate the corresponding coupling
constants. We assume that the following resonance conditions are
satisfied:
\begin{equation}
E_4 - E_1 = 3 \omega_b ,
\qquad
E_3 - E_1 = 2 \omega_b ,
\end{equation}
while all one-photon transitions are out of resonance. In
such a case, the Hamiltonian (\ref{H4}) may describe, for example,
a fifth-order process involving the absorption of three photons
from one field and the stimulated emission of two photons of
different frequency. The effective Hamiltonian describing
\textit{explicitly} these transitions can be obtained according
to the general method. To this end, we note that now
the integral of motion (\ref{IM}) takes now the form
\begin{equation}
N = a^\dagger a + b^\dagger b + \sum_{i=1}^{3} \mu_i S_3^{i i+1},
\end{equation}
where $\mu_i = i (4-i)$, and the Hamiltonian can be recast
also as in (\ref{recast}) with $H_{\mathrm{int}} = h_0 + V_a + V_b$,
where
\begin{eqnarray}
& \displaystyle h_0  =  \delta \ b^\dagger b +
\sum_{i=1}^{4} \Delta_i S^{ii} , & \nonumber \\
& & \\
& \displaystyle V_a   =    \sum_{i=1}^{3} g_{ai} (X_{a+}^{i i+1} + X_{a-}^{i i+1})
\qquad
 V_b   =    \sum_{i=1}^{3} g_{bi} (X_{b+}^{i i+1} + X_{b-}^{i i+1}) . &
\nonumber
\end{eqnarray}
Here the polynomial deformation at hand is defined by
\begin{equation}
X_{a+}^{ij}= a S_+^{ij},
\quad
X_{a-}^{ij}= a^\dagger S_-^{ij} ;
\qquad
X_{b+}^{ij}= b S_+^{ij},
\quad
X_{b-}^{ij}= b^\dagger S_-^{ij} ,
\end{equation}
and $\delta = \omega_b - \omega_a$ will be taken, for definiteness, as positive.

Now we can eliminate all the one-photon nonresonant transitions generated
by means of a couple of transformations (one for mode $a$ the other for
mode $b$) identical to (\ref{T1}). The transformed Hamiltonian (up to order
$1/\Delta^2$) is
\begin{eqnarray}
\label{H4ef}
H_{\mathrm{eff}}^{(1)}  & = & h_0 +
h_{\mathrm{d}}^{(a)} + h_{\mathrm{d}}^{(b)} \nonumber \\
& + & \frac{1}{2} \lambda_1^{(2)}
\left ( a^2 S_+^{13}+ {a^\dagger}^2 S_-^{13} \right )
+ \frac{1}{3} \lambda_1^{(3)}
\left ( b^3 S_+^{14}+ {b^\dagger}^3 S_-^{14} \right ) +
\xi_2^{(ab)}
\left ( a b S_+^{24}+ a^\dagger b^\dagger S_-^{24} \right ) ,
\end{eqnarray}
where $h_{\mathrm{d}}^{(a)}$ and $h_{\mathrm{d}}^{(b)}$ are
defined according to  (\ref{diag}) for both modes and the coupling
constants $\lambda_1^{(2)}$ and $\lambda_1^{(3)}$ are given by
(\ref{lacou}).  The constant $\xi_2^{(ab)}$ is
\begin{equation}
\xi_2^{(ab)} = \frac{g_{a3}g_{b2}}{\Delta_4 -  \Delta_3} -
 \frac{g_{b3}g_{a2}}{\Delta_3 -  \Delta_2} .
\end{equation}

This effective Hamiltonian deserves some comments. First of all,
when the resonance condition
\begin{equation}
E_4 - E_2 = \omega_a + \omega_b
\end{equation}
is fulfilled (which is compatible with the two- and three-photon resonance
conditions), then the last term in the Hamiltonian describes resonant
transitions between levels 2 and 4, as a  result of the simultaneous
absorption and emission of quanta from both modes, and take place
only if levels 2 and 4 are populated. In absence of such an additional
resonance condition, the  Hamiltonian (\ref{H4ef}) describes two
simultaneous and competing processes: a first-order process ($\lambda_1^{(2)}
\sim 1/\Delta$) involving two-photon transitions between levels 1 and 3, and
a second-order process ($\lambda_1^{(3)} \sim 1/\Delta^2$) of
three-photon transitions between levels 1 and 4.

\section{Further extensions and concluding remarks}

We have developed our method putting special emphasis
in polynomial deformations of algebras su($N$), due to the
outstanding role that they play in describing the interaction
of $N$-level systems with quantum fields. However,
by no means the method is restricted to such kind of algebras.
In fact, many other phenomena can be modeled
by Hamiltonians quite similar to (\ref{Hint}), namely
\begin{equation}
H_{\mathrm{int}} = a \ Y_0 + g (Y_+ + Y_-) + C ,
\end{equation}
where $C$ is some integral of motion and $a$ a constant.
In these theories, the polynomial deformation is defined in the
following fashion in the Cartan-Weyl basis:
\begin{equation}
[Y_0 , Y_\pm] = \pm Y_\pm ,
\qquad
[Y_- , Y_+] = \Psi (Y_0) = \Phi(Y_0 + 1) - \Phi(Y_0)  ,
\end{equation}
where $\Phi(Y_0)$ are appropriate structure polynomials.
A detailed study of the applications of these algebras
for solving evolution problems in nonlinear quantum
models may be found in Ref. \cite{Karasiov94}. For
these model, the machinery of small rotations works well
and constitutes the most systematic ways of constructing
effective Hamiltonians.

In summary, what we expect to have accomplished
in this paper is to develop an appropriate and systematic
tool for obtaining effective Hamiltonians that describe
nonlinear optical phenomena. Our approach exploits the
existence of deformed algebras (arising as dynamical
symmetries of the corresponding process) to construct
small nonlinear rotations that perform the task.

This method constitutes a firm algebraic setting free
from some inconsistencies found in the traditional adiabatic
elimination of variables and provides unambiguous recipes
for proceeding with any model Hamiltonian.


\begin{thebibliography}{99}

\bibitem{Cohen98}
Cohen-Tannoudji, C., Dupont-Roc, J., and Grynberg G., 1998,
\textit{Atom-photon interactions: basic processes and applications}
(New York: Wiley).

\bibitem{Walls95}
Walls, D. F., and  Milburn, G. J., 1995,
\textit{Quantum optics}
(Berlin: Springer).

\bibitem{Scully99}
Scully, M. O., and  Zubairy, M. S., 1999,
\textit{Quantum optics}
(Cambridge: Cambridge University Press).

\bibitem{Perina91}
Pe\v{r}ina, J., 1991,
\textit{Quantum statistics of linear and
nonlinear optical  phenomena}
(Dordrecht: Kluwer).

\bibitem{Schenzle82}
Schenzle, A., 1981,
\textit{Nonlinear optical phenomena and fluctuations},
Lecture Notes in Physics \textbf{155}
(Berlin: Springer) pg. 103.

\bibitem{Graham68a}
Graham, R.,  and Haken, H., 1968,
\textit{Z. Phys.}  \textbf{210}, 276-291.

\bibitem{Graham68b}
Graham, R., 1968,
\textit{Z. Phys.} \textbf{210}, 319.

\bibitem{Bloembergen92}
Bloembergen, N., 1990,
\textit{Nonlinear optics}
(New York: John Wiley).

\bibitem{Shen85}
Shen, Y. R., 1985,
\textit{The principles of nonlinear optics}
(New York: John Wiley).

\bibitem{Puri88}
Puri, R. R., and Bullough, R. K., 1988,
\textit{J. Opt. Soc. Am. B} \textbf{5}, 2021-2028.

\bibitem{Gigi90}
Lugiato, L. A.,  Galatola, P., and  Narducci, L. M., 1990,
\textit{Opt. Commun.} \textbf{76}, 276-286.

\bibitem{KC}
Klimov, A. B., Negro, J., Farias, R., and  Chumakov, S. M., 1999,
\textit{J. Opt. B: Quantum Semiclass. Opt.} \textbf{1}, 562-570.

\bibitem{Sczaniecki83}
Sczaniecki, L., 1983,
\textit{Phys. Rev. A} \textbf{28}, 3493-3514.

\bibitem{Hillery85}
Hillery, M., and Mlodinow, L. D., 1985,
\textit{Phys. Rev. A} \textbf{31}, 797-806.

\bibitem{Klein74}
Klein, D. J., 1974,
\textit{J. Chem. Phys.} \textbf{61}, 786-798.

\bibitem{Kittel87}
Kittel, C., 1987, \textit{Quantum theory of solids}
(New York: John Wiley) pp. 148-151.

\bibitem{Shavitt80}
Shavitt, I., and  Redmon, L. T., 1980,
\textit{J. Chem. Phys.} \textbf{73}, 5711-5717.

\bibitem{Karasiov92}
Karassiov, V. P., 1992,
\textit{J. Sov. Laser Research} \textbf{13}, 188-195;
Karassiov, V. P., 1994,
\textit{J. Phys. A} \textbf{27}, 153-165.

\bibitem{Karasiov94}
Karassiov, V. P.,  and Klimov, A. B., 1994,
\textit{Phys. Lett. A} \textbf{189}, 43-51.

\bibitem{Debergh97}
Debergh, N., 1997,
\textit{J. Phys. A} \textbf{30}, 5239-5242.

\bibitem{Delgado00}
Delgado, J., Luis, A.,  S\'anchez-Soto, L. L.,  and Klimov, A. B., 2000,
\textit{J. Opt. B: Quantum Semiclass. Opt.} \textbf{2}, 33-40.

\bibitem{Sunil00}
Sunilkumar, V.,  Bambah, B. A.,  Jagannathan, R.,
Panigrahi, P. K.,  and Srinavasan, V., 2000,
\textit{J. Opt. B: Quantum Semiclass. Opt.} \textbf{2}, 126-132.

\bibitem{Klimov00}
Klimov, A. B.,  and  S\'anchez-Soto, L. L., 2000,
\textit{Phys. Rev. A} \textbf{61}, 063802 1/11.

\bibitem{Steinberg87}
Steinberg, S., 1987,
\textit{Lie methods in optics},
Lecture Notes in Physics \textbf{250}
(Berlin: Springer), pg. 45.

\bibitem{Taka75}
Takarsuji, M., 1975,
\textit{Phys. Rev. A} \textbf{11}, 619-624.

\bibitem{Loy75}
Loy, M. M. T.,  and  Liao, P. F., 1975,
\textit{Phys. Rev. A}, \textbf{12}, 2514-2533.

\bibitem{Stroud76}
Whitley, R. M.,  Stroud jr., C. R., 1976,
\textit{Phys. Rev. A} \textbf{14}, 1498-1513.

\bibitem{Cohen77}
Cohen-Tannoudji, C.,  and Haroche, S., 1977,
\textit{J. Phys. B} \textbf{10}, 365-384.

\bibitem{BO78}
Bowden, C. M., and  Sung, C. C., 1978,
\textit{Phys. Rev. A} \textbf{18}, 1558-1570.

\bibitem{Mossberg77}
Mossberg, T. W., Flusberg, A.,  Kachru, R.,  and  Hartmann, S. R.,  1977,
\textit{Phys. Rev. Lett.} \textbf{39}, 1523-1526.

\bibitem{Mossberg81}
Mossberg, T. W., and   Hartmann,  S. R., 1981,
Phys. Rev. A \textbf{23}, 1271-1280.

\bibitem{Perinova94}
Pe\v{r}inova, V.,  and  Luk\v{s}, A., 1994,
in \textit{Progress in Optics}, edited by
E. Wolf   (Amsterdam: North-Holland), vol. 33.

\bibitem{Cohen90}
Cohen-Tannoudji, C., Diu, B., and Lalo\"{e}, F., 1992,
\textit{Quantum mechanics}
(New York: Addison-Wesley).

\bibitem{Skl82}
Sklyanin, E. K., 1982,
Funct. Anal. Appl. \textbf{16}, 263-273.

\bibitem{Higgs79}
Higgs, P. W., 1979,
\textit{J. Phys. A} \textbf{12}, 309-323.

\bibitem{Rocek91}
Ro\^{c}ek, M., 1991,
\textit{Phys. Lett. B} \textbf{255}, 554-557.

\bibitem{Bonatsos93}
Bonatsos, D.,  Daskaloyannis, C.,  and  Lalazissis, G. A., 1993,
\textit{Phys. Rev. A} \textbf{47}, 3448-3451.

\bibitem{Quesne95}
Quesne, C., 1995,
\textit{J. Phys. A} \textbf{28}, 2847-2860.

\bibitem{Beckers96}
Abdesselam, B.,  Beckers, J.,  Chakrabart, A.,  and  Debergh, N., 1996,
\textit{J. Phys. A} \textbf{29}, 3075-3088.

\bibitem{Dicke54}
Dicke, R., 1954,
\textit{Phys. Rev.} \textbf{93}, 99-110.

\bibitem{Brune96}
Brune, M., Hagley, E.,  Dreyer, J.,
Ma\^{\i}tre, X.,  Maali, A.,  Wunderlich, C.,
Raimond, J. M.,  and  Haroche, S., 1996,
\textit{Phys. Rev. Lett.} {\bf 77}, 4887-4890.

\bibitem{Agarwal97}
Agarwal, G. S., Puri, R. R., and  Singh, R. P., 1997,
\textit{Phys. Rev. A} \textbf{56}, 2249-2254.

\bibitem{Klimov98}
Klimov, A. B.,  and  Saavedra, C., 1998.
\textit{Phys. Lett. A} \textbf{247}, 14-20.

\bibitem{Ueda93}
Kitagawa, M.,  and Ueda, M., 1993,
\textit{Phys. Rev. A} \textbf{47}, 5138-5143.

\bibitem{YE85}
Yoo, H. I.,  and  Eberly, J. H., 1985,
\textit{Phys. Rep.} \textbf{118}, 239-337.



\end{thebibliography}
\end{document}